# Yield Strength-Plasticity Trade-off and Uncertainty Quantification for Machine-learning-based Design of Refractory High-Entropy Alloys

Stephen A. Giles[1], Hugh Shortt[2], Peter K. Liaw[2], Debasis Sengupta[1*]

[1] CFD Research Corporation, 6820 Moquin Drive NW, Huntsville, AL 35806

[2] Department of Materials Science and Engineering, The University of Tennessee, Knoxville, TN 37996.

**Abstract**

Development of process-structure-property relationships in materials science is an important and challenging frontier which promises improved materials and reduced time and cost in production. Refractory high entropy alloys (RHEAs) are a class of materials that are capable of excellent high-temperature properties. However, due to their multi-component nature, RHEAs have a vast composition space which presents challenges for traditional experimental exploration. Here, quantitative models of compressive yield strength and room-temperature plasticity are developed through a deep learning approach. Uncertainty quantification is performed through a variety of statistical validation techniques. Model predictions are experimentally validated through collection of recent literature and the synthesis and experimental characterization of two new, unreported RHEAs: AlMoTaTiZr and $Al_{0.239}Mo_{0.123}Ta_{0.095}Ti_{0.342}Zr_{0.201}$. Finally, through the application of model interpretability, features having the greatest impact on both the mechanical property and uncertainty of the deep learning models are revealed, and shown to agree well with current physics and materials science theory.

*Correspondence: debasis.sengupta@cfd-research.com

## Introduction

High entropy alloys (HEAs) are an emerging class of materials which diverge from traditional alloys by containing multiple (typically, ≥ 4) elements in near-equal fractions, exploiting high configurational entropy to help drive their thermodynamic stability[1,2]. Since their discovery in 2004, more than 1500 HEAs have been synthesized and characterized, leading to the discovery of a number of HEAs which have exhibited superior thermomechanical properties[3–9]. In particular, refractory HEAs (RHEAs) have the potential for superior performance in high-temperature applications, such as high-efficiency gas turbines. However, while compressive strengths of RHEAs at high temperatures are reported to outperform those of Ni superalloys, their room-temperature ductilities must be improved significantly. For mechanical properties, it has been widely established that improvement of strength comes at the cost of room-temperature ductility (or plasticity in the case of compressive testing). Therefore, identifying alloy composition with trade-off properties is one of the major challenges in this area.

To aid in the rational design of RHEAs, traditional or modified microstructural-based alloy theories have been used to rationalize experimental observations. For example, edge[10] and screw[11] dislocation theories have been used for explaining and designing RHEAs with high-temperature yield strength. Density functional theory (DFT) is also used quite frequently for obtaining fundamental understanding of experimental observations. With respect to plasticity, the "D" parameter, which is the ratio of surface energy to unstable stacking fault energy of cleavage planes calculated using DFT, has been correlated with experimental measures of plasticity[12]. Wu *et al.* have recently demonstrated a theoretical basis for the origin of plasticity in magnesium-containing alloys through consideration of the energetic barrier for screw dislocation cross-slip[13]. More recently, property prediction models with machine learning (ML) methods have gained considerable attention as an alternative solution to high-fidelity DFT modeling and experimentation. We have recently reported an ML model for prediction of compressive yield strength with a limited number of RHEA data from Couzinie *et al.*, and also used the model to develop a framework for discovering new RHEA composition through optimization[14]. While ML methods have immense potential for rapid exploration of compositional space, it often suffers from a lack of physical interpretability. Moreover, little attention has been given on the uncertainty of the ML model predictions, since generating statistically significant validation data through alloy synthesis and property determination is tedious and time-consuming. Experimental uncertainties of measured properties are also not reported, and variation in the measured properties among various research groups is rather common.

Here, we have refined this model via training with a recently published much larger dataset using various fully-connected neural networks. We also developed a ML model for plasticity prediction using the data derived from the stress-strain plots of alloys reported in the references of Couzinie database. Comprehensive analysis of model performance via error analysis using various validation methods. Such an analysis provides a quantitative estimation of model uncertainty. The yield strength and plasticity model predictions are further validated using unseen recent literature data, as well as our own experimental data of two new alloys, AlMoTaTiZr and $Al_{0.239}Mo_{0.123}Ta_{0.095}Ti_{0.342}Zr_{0.201}$. Finally, model interpretation is performed using state-of-the-art model explainability techniques. Our analysis is not only in agreement with well-established physics, but also provides additional insights that could lead to important physics for future study. The present work enables rapid exploration through high-throughput predictions and simultaneous optimization of alloy properties.

**Results**

The general workflow of data ingestion, model training, validation, interpreting, and new alloy prediction is depicted in Figure 1a. As part of the data ingestion process in Figure 1a, the data was filtered to include only *compressive* yield strength and *room-temperature* plasticity. Inclusion of only compression data was done due to the different physics of mechanical strain under compression versus tension, and the comparatively less data available for tension properties. Similarly, room-temperature plasticity is the focus of this work since room temperature is the most common condition tested, and that the brittleness which limits some alloys from practical use is most problematic in the low-temperature regime. Two primary data sources were used for model development. The RHEA yield strength data were obtained from the recently published multi-principal element alloy (MPEA) database by Citrine Informatics[15]. This database includes both RHEAs and 3*d* transition metal high entropy alloys. According to the standard definition of refractory elements (Mo, Nb, Re, Ta, and W), 791 of the 1524 MPEA compositions (52%) can be classified as RHEAs. Of these 791 compositions, 525 have reported compressive yield strength data. In contrast to yield strength, data on plasticity (i.e., plastic strain at fracture) are not readily available. Thus, we have curated our own RHEA plasticity dataset by extracting data from the original references cited by Couzinie *et al.* in their review on RHEAs[16]. The dataset we have collected is comprised of a total of 257 measurements, of which, 116 measurements are at room temperature and constitute the dataset leveraged here. We refer readers to the Supplementary Information (SI) where the datasets of both yield strength and plasticity are provided. On the basis of the compositions contained in the yield strength and plasticity datasets, a set of composition-based features were generated using Matminer[17]. More details are provided in the Methods Section.

*Model Development and Statistical Validation*

The feature sets created from the yield strength and plasticity datasets were used to perform model training. We first focused on validating a single model trained on the full yield strength and plasticity datasets, respectively. To assess generalizability of this model, repeated *k*-fold cross-validation was performed. Three-fold, five-fold, and ten-fold cross-validation were performed in order to assess trends with increasing training set size. Additionally, leave-one-out cross-validation (LOOCV) was also performed and compared to the *k*-fold cross-validation results as a limiting case where all but one data point are used for training. As part of the optimization of the neural network structure, we have compared the performance of three different activation functions: rectified linear unit (ReLU)[18], Leaky ReLU[19], and scaled exponential linear unit (SELU)[20]. Additionally, random forest and gradient boosting models were each constructed. A comparison of the model validation uncertainty estimates of the compressive yield strength and room-temperature plasticity models is provided in Figure 1b and Figure 1c, respectively. The RMSE distributions for three-fold, five-fold, and ten-fold cross-validation using Leaky ReLU activation, the best performing activation function according to *k*-fold cross-validation, are provided in Figure 1d and Figure 1e for compressive yield strength and room-temperature plasticity, respectively. In regards to the RMSE distributions, corresponding distributions determined for all other models are provided in the Figures S1-S8 for both yield strength and plasticity. When moving from 3-fold to 10-fold cross-validation, the RMSE is generally observed to decrease. This effect is a result of the growing training set size. The same phenomenon has been in observed in other studies which compare the error estimates determined from different validation schemes[21]. Each training iteration of 3-fold, 5-fold, and 10-fold cross-validation is analogous to 33% hold-out, 20% hold-out, and 10% hold-out. Therefore, we directly compared the RMSE of hold-out validation to the corresponding *k*-fold cross-validation and found that they are in quantitative agreement, as expected (Figure S9). Comparing the error determined from 10-fold cross-validation and LOOCV, it is important to note that LOOCV has no standard deviation in the RMSE due to its having a single, unique training dataset, consisting of every *other* point in the dataset. The error distribution of individual test data points, however, has a wide distribution with a standard deviation equal to the computed RMSE (Figure S10).

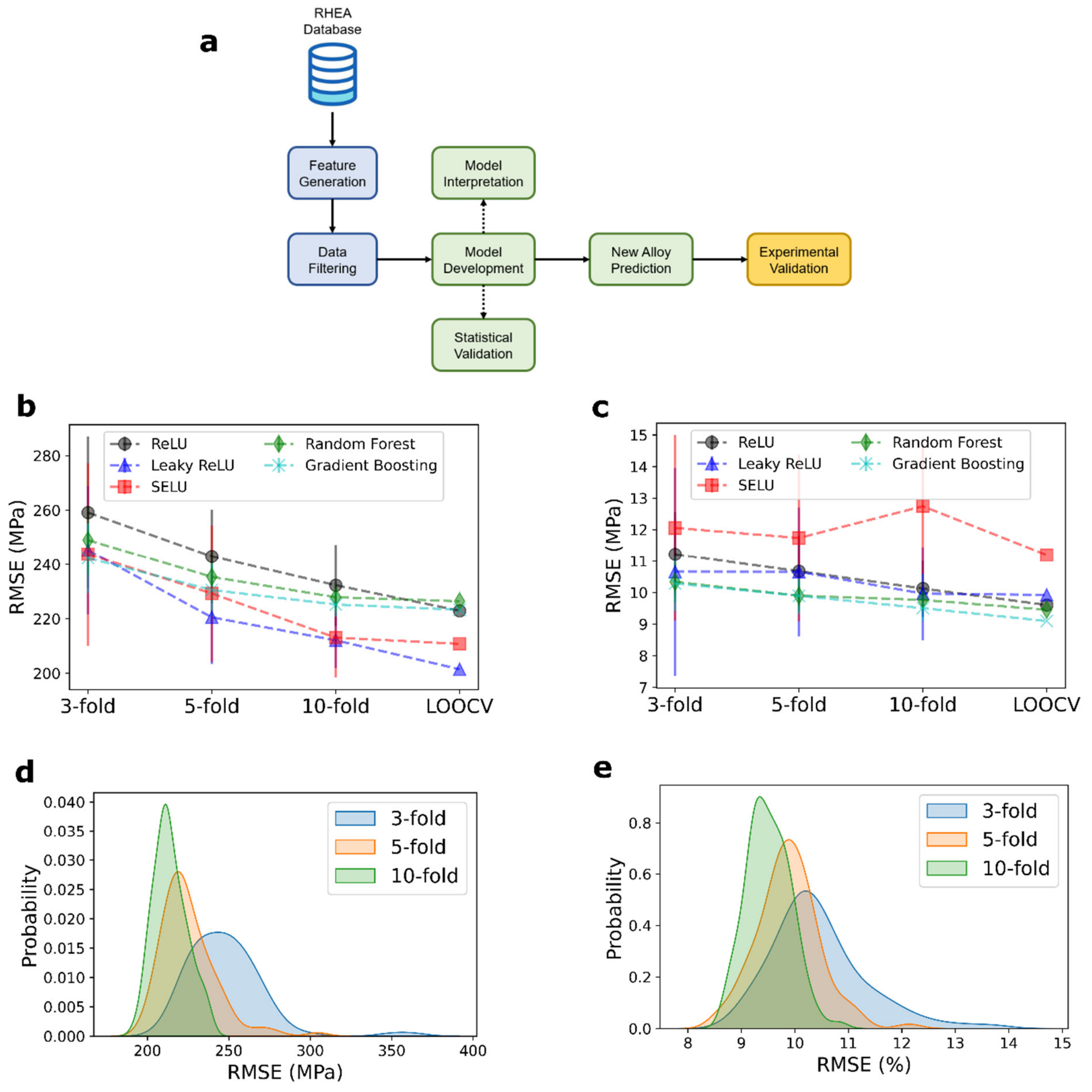

**Figure 1.** (a) Workflow for developing, validating, and interpreting deep learning models for RHEA property prediction. (b) 3-fold, 5-fold, 10-fold, and leave-one-out cross-validation

(LOOCV) error of yield strength for each of the three activation functions under consideration. The mean and standard deviation across 100 *k*-fold repetitions are shown for the 3-, 5-, and 10-fold results. (c) 3-fold, 5-fold, 10-fold, and leave-one-out cross-validation (LOOCV) error of room-temperature plasticity for each of the three activation functions under consideration. The mean and standard deviation across 100 *k*-fold repetitions are shown for the 3-, 5-, and 10-fold results. (d) Distribution of error in yield strength prediction, depicted by kernel density estimate (KDE), for the Leaky ReLU activation as a function of number of folds. (e) Distribution of error in room-temperature plasticity prediction, depicted by KDE, for the Leaky ReLU activation as a function of number of folds.

The distribution of RMSE values for the cross-validation study represents a characteristic error for a given model iteration (i.e., a single repetition of 3-, 5-, or 10-fold CV). Consequently, the RMSE distribution corresponds to the expected error when a *single* model is trained on the entire dataset and used for inference. However, it has been observed that *ensemble* models are often capable of outperforming singular models, in addition to providing estimates of prediction-specific uncertainty[22–25]. In this respect, bootstrapping is arguably the most widely used ensemble approach where individual models are created through repeated sampling (with replacement) of the dataset and training a model on each sample[26]. The connection between bootstrapping and the cross-validation study is that if, for each data point, the test prediction value (i.e., the value predicted when the data point is contained in the test fold) is averaged across all *k*-fold repetitions, then the resulting mean prediction is analogous to the uncertainty of a bootstrap ensemble model. Therefore, analyzing the cross-validation results in this manner, we discover that averaging the predictions of each data point leads to a decreased error, as indicated by the RMSE in the yield strength parity plot (Figure 2a). The RMSE of 187.1 MPa is a 7.1% reduction in the median RMSE of 201.5 MPa for the corresponding 10-fold cross-validation with Leaky ReLU activation in Figure 1. An even greater benefit is also observed for the room-temperature plasticity (Figure 2c), where the plasticity RMSE decreased from 9.96% to 8.38%, a relative reduction of 15.9%. In Figure 2b and 2d, the distributions of the standard deviation in the *k*-fold yield strength and plasticity predictions are provided. These distributions are related to the model uncertainty (i.e., epistemic uncertainty). Provided alongside the distribution of epistemic uncertainties in Figure 2b and Figure 2d, we include uncertainty in the yield strength and plasticity data (i.e., aleatoric uncertainty) to understand which has the greatest impact on the calculated RMSEs of the models. Specifically, within the Citrine MPEA dataset, there are a number of examples of independently reported, duplicate measurements for the same alloy composition and experimental conditions. However, due to random experimental error, as well as potential differences in experimental conditions that are not present in the dataset (e.g., annealing temperature, cooling rate, etc.), the alloy properties differ for theoretically equivocal measurements. To approximate the aleatoric uncertainty in the yield strength, we have identified all measurements within the Citrine dataset that were independently reported by more than one source and calculated their standard deviation (mean and standard deviation of each replicated set of experimental conditions is provided in the SI). The aleatoric uncertainty is found to have a median value of 148.2 MPa, twice that of the epistemic uncertainty. Therefore, the bulk of the error in the models presented here has its origin in uncertainty in the experimental data, and not in insufficiency of the models. As summarized in Figure 2e, aggregation of predictions through a bootstrap is found to reduce the total error of both the yield strength and plasticity models, in addition to its inherent quantification of epistemic uncertainty.

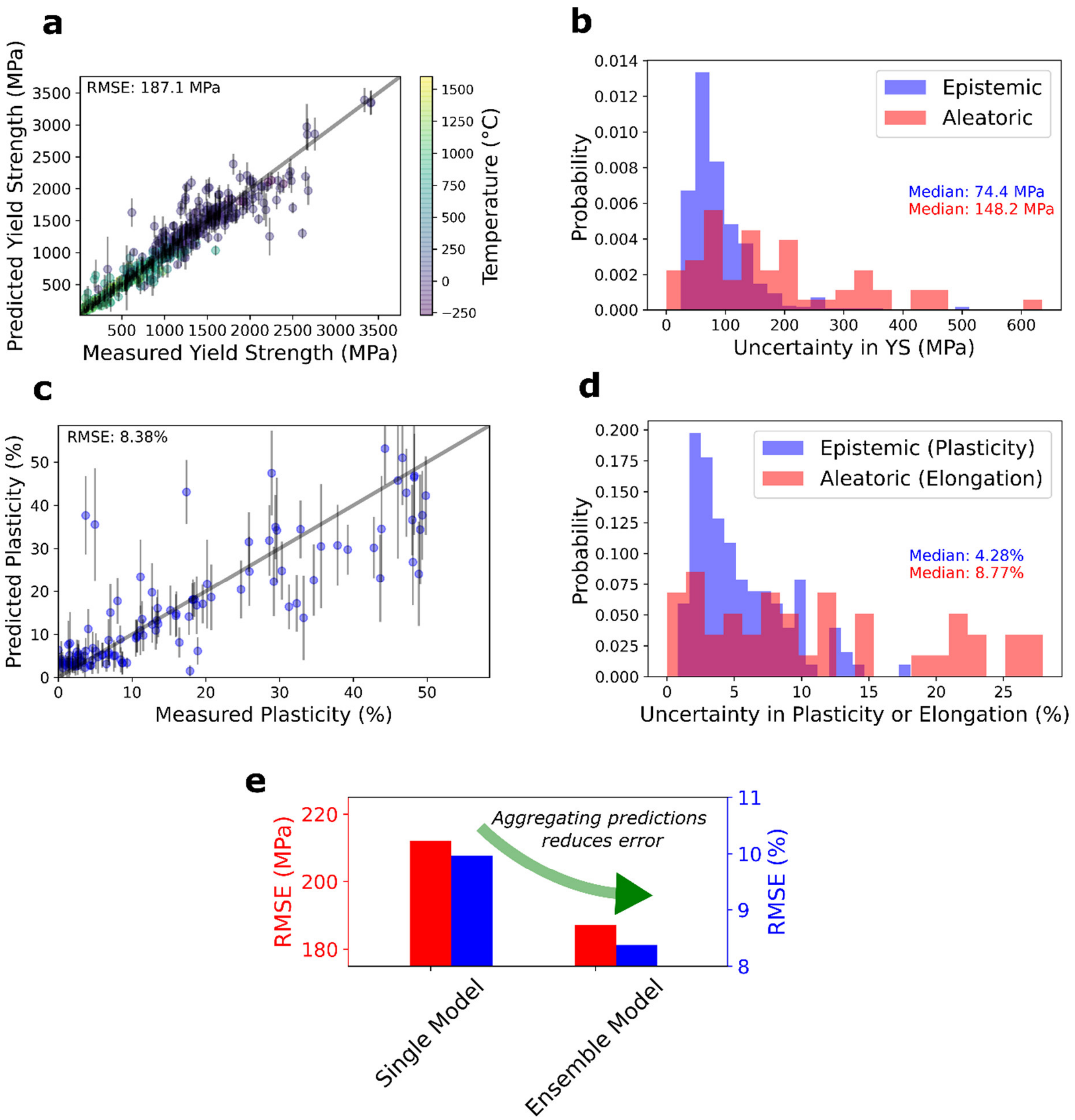

**Figure 2.** Summary of prediction distribution obtained from averaging predictions across all *k*-fold replicates, analogous to bootstrapping. (a) Parity plot of 10-fold cross-validation using Leaky ReLU activation. Data points are colored according to the test temperature. The mean and standard deviation across 100 *k*-fold repetitions are shown. (b) Distribution of standard deviation in *k*-fold replicates for the 10-fold cross-validation using Leaky ReLU activation (epistemic uncertainty) shown with uncertainty distribution in experimental measurements from replicated data in the Citrine database[15]. (c) RMSE of Leaky ReLU plasticity model. The mean and standard deviation across 100 *k*-fold repetitions are shown. (d) Distribution of standard deviation in plasticity in *k*-fold replicates for the 10-fold cross-validation using Leaky ReLU activation (epistemic uncertainty) shown with uncertainty distribution in experimental measurements from replicated data in the Citrine database[15]. (e) RMSE of yield strength and plasticity models determined from averaging the individual test predictions in 3-, 5-, and 10-fold cross-validation for each activation function.

*Validation with Recent Literature Data*

A significant number of recently published RHEA data on compressive yield strength and room-temperature plasticity not included in either the Citrine MPEA database or the Couzinie *et al.* dataset was collected to serve as a statistically significant validation set completely unseen to the models developed here. These totaled 178 compressive RHEA yield strength measurements and 59 room-temperature RHEA plasticity measurements. The RHEA compositions collected were all compositions which did not appear in the training set. Parity plots of the yield strength and plasticity are provided in Figure 3a and Figure 3b. A summary of the RMSEs are provided in Figure 3c. The results are in general agreement with the *k*-fold cross-validation results, although the errors for the unseen validation set are somewhat larger. This discrepancy primarily results from the existence of multiple measurements independently reported for the same compositions. As observed in Figure S11, when considering only *unique* alloy compositions within the dataset, the errors determined from *k*-fold cross-validation approach in Figure 2 are in close agreement with the error for the unseen validation set. Following use of the additional data points to validate the models, the newly collected yield strength and plasticity data were included in the training dataset, the models were retrained, and the error was measured through 10-fold cross-validation. Inclusion of the additional data resulted in insignificant changes to the yield strength and plasticity RMSEs, as determined from 10-fold cross-validation. Because all the additional data used to augment the dataset were *unique* compositions, the retrained models are expected to have greater generalizability to future unseen data.

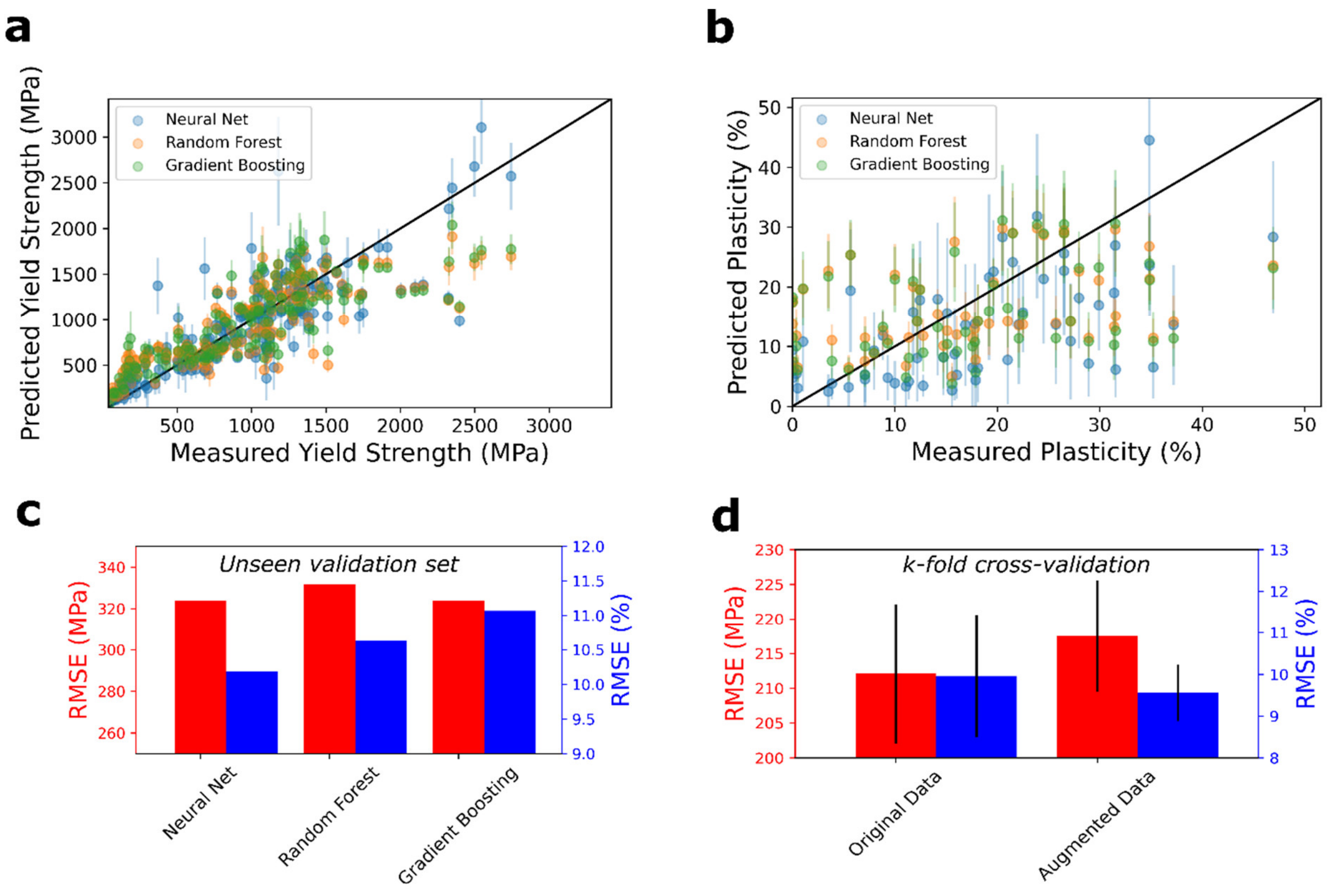

**Figure 3.** Validation and re-training of the yield strength and plasticity models using unseen recent literature data. (a) Parity plot of measured yield strength versus predicted yield strength for the Neural Network (Leaky ReLU), Random Forest, and Gradient Boosting ensemble models. (b) Parity plot of measured plasticity versus predicted plasticity for the Neural Network (Leaky ReLU), Random Forest, and Gradient Boosting ensemble models. (c) Summary of RMSEs on the unseen recent literature validation set. (d) Comparison of the RMSEs calculated through repeated 10-fold cross-validation for the original yield strength and plasticity models, and the re-trained models with on the augmented dataset including the recent literature data from parts (a) and (b). Error bars depict the standard deviation of the RMSE amongst the repeated 10-fold validations.

*Experimental Validation*

In a recent paper, we have previously identified AlMoTaTiZr, a previously unreported base alloy, and $Al_{0.239}Mo_{0.123}Ta_{0.095}Ti_{0.342}Zr_{0.201}$ (hereafter referred to as "modified AlMoTaTiZr") as being particularly strong alloys[14]. The experimental stress-strain curves are provided in Figure 4a. Figure 4b and 4c show the neural network ensemble model predictions for compressive yield strength and room-temperature plasticity, respectively. The experimentally measured yield strengths of both AlMoTaTiZr and modified AlMoTaTiZr are shown to lie within the uncertainty of the model prediction. The compressive yield strength model predicts the correct qualitative trend with respect to the equiatomic AlMoTaTiZr alloy having the higher yield strength. Between the two alloys, the average error of the ensemble model is 197 MPa, which is in close agreement with the error estimated from cross-validation. Likewise, the experimentally determined plasticity of both AlMoTaTiZr and modified AlMoTaTiZr are shown in Figure 4b to agree very well with the plasticity ensemble model. Both alloys were experimentally determined to be exceptionally brittle (plasticity < 1%), and the ensemble model correctly identified this brittle behavior with plasticity predictions of 3.2% and 4.3% for AlMoTaTiZr and modified AlMoTaTiZr, respectively. Moreover, the experimentally determined plasticities of both alloys are within the statistical distribution of the model predictions. Thus, while within the scope of the present work, a limited amount of *experimental* validation could be performed, the *statistical* validation strategy adopted herein allows for each alloy in the dataset of 525 compressive yield strength measurements and in the dataset of 117 room-temperature plasticity measurements to be treated as validation data, thus giving a far more universal understanding of the model error than could feasibly be attained through traditional experimental validation.

The temperature-dependent yield strength predictions of the ensemble neural network model are shown in Figure 4d and 4e for AlMoTaTiZr and modified AlMoTaTiZr, respectively. Also noteworthy is that AlMoTaTiZr is shown to have a particularly high yield strength of (872 ± 199 MPa) at 1,000 °C. A yield strength of 872 MPa at 1,000 °C would be the strongest *as-cast* HEA discovered to date, according to the experimental data contained in the Citrine database and our most recent knowledge. The AlMoTaTiZr material is most similar to a previously reported, high-performing alloy, HfMoTaTiZr, which has an as-cast yield strength of 855 MPa at 1,000 °C[27]. The AlMoTaTiZr material reported here, however, has the advantage of being lighter weight (rule-of-mixtures density of 8.1 g·cm$^{-3}$ compared to 10.3 g·cm$^{-3}$) and lower in melting temperature (rule-of-mixtures melting temperature of 1964 °C compared to 2279 °C), thus making for more facile experimental synthesis.

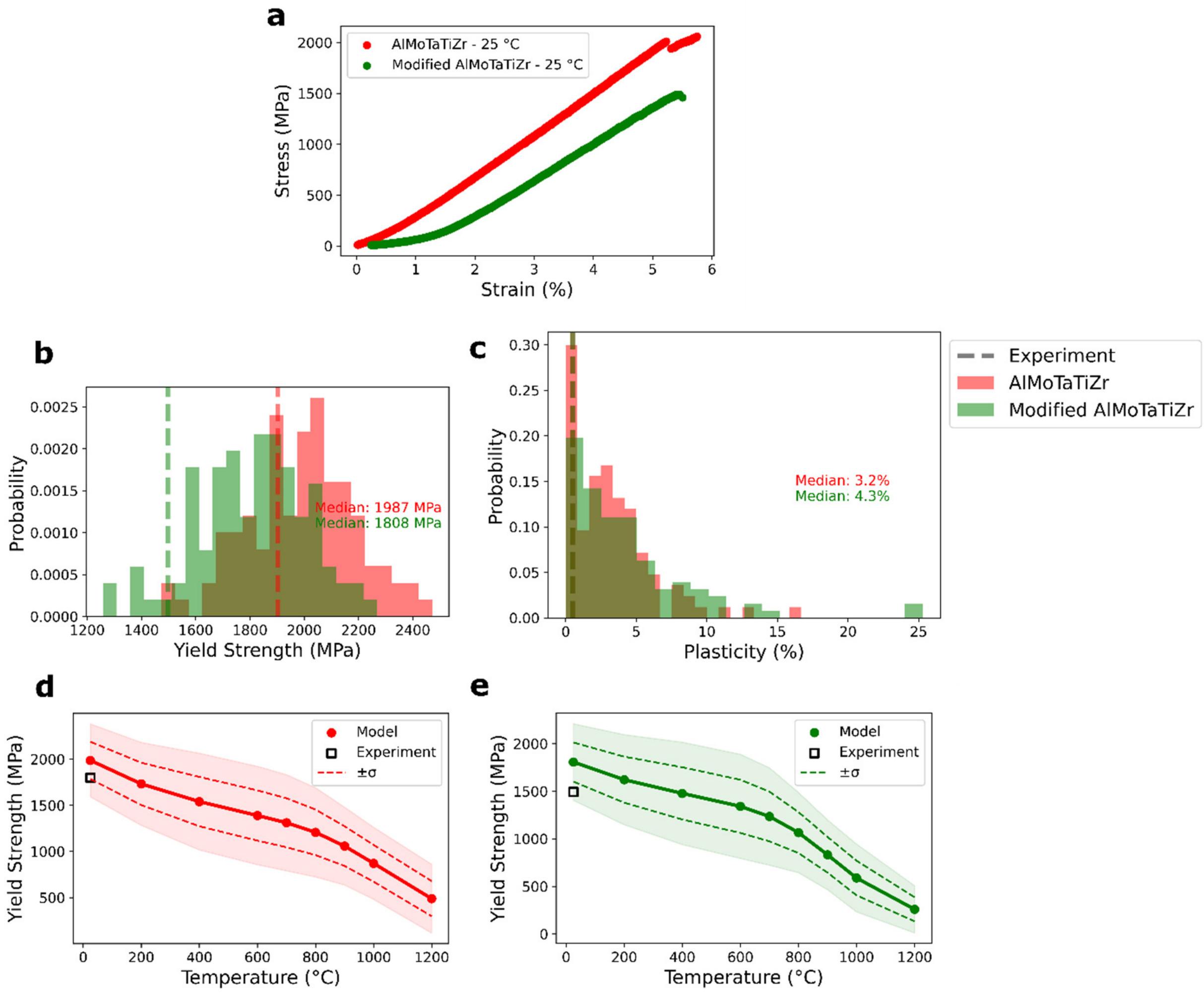

**Figure 4.** Experimental validation of the yield strength and plasticity RHEA models. (a) Representative stress-strain responses at room temperature for AlMoTaTiZr and modified AlMoTaTiZr. Experimental values provided for yield strength and plasticity were averaged among multiple repeated sample measurements. Neural network ensemble model predictions of yield strength (b) and plasticity (c) for AlMoTaTiZr (shown in red) and modified AlMoTaTiZr (shown in green). Median values from the ensemble model are shown. The experimentally measured value for both materials is shown with a vertical dashed line. Temperature-dependent yield strength predictions of the neural network ensemble for AlMoTaTiZr (d) and modified AlMoTaTiZr (e). The experimentally measured value (same as those shown in part (b) with the vertical dashed lines) are indicated with the black square. The solid line, dashed line, and shaded region represent the mean prediction, ± one standard deviation, and the 95% prediction interval, respectively.

*Model Interpretation*

Validation of the physical implications of data-driven models is an important undertaking since minimizing the uncertainty of these models has been demonstrated to sometimes lead to a lack of interpretability and physical insights, even for specifically predicting alloy properties[28]. To interpret the models discussed here, we perform Shapley Additive Explanations (SHAP) analysis, a suitable framework for quantitatively determining feature effects within a deep learning model[29–33]. A summary of the SHAP values for the ten most important features is provided for both yield strength and plasticity in Figure 5a and Figure 5b, respectively. For the yield strength ensemble model, the most important feature was confirmed to be the test temperature, which is unsurprising and ubiquitously important. Other composition-dependent features, however, are shown to have a significant impact on the yield strength and contribute to differentiating different alloys from one another. Among the ten most important features, six of the features are elemental fractions: Zr, Cr, Mo, V, Ni, and Ti. Zr, Cr, Mo, and V were found to be positively correlated with yield strength, whereas Ni and Ti were found to be negatively correlated. Physics-based descriptors, such as the modulus distortion of Ta ($\delta G_{\mathrm{Ta}}$), were also found to positively affect the yield strength. A similar observation was made for the key features affecting the plasticity, with $\delta r_{\mathrm{Hf}}$, $\delta G_{\mathrm{Mo}}$, and $\delta r_{\mathrm{V}}$ being among the most important features. In addition, the annealing temperature was found to have a strong negative impact on the plasticity, a frequently observed experimental phenomenon.

Given that SHAP values were calculated for each member of the ensemble model, not only can the mean SHAP values be calculated for every data point, but also the standard deviation. Consequently, for each data point, the variance of feature effect on the model output is known and can be used to trace back origins of uncertainty within the model. The contribution of individual features to the total observed variance in the SHAP values is depicted for the yield strength and plasticity models in Figure 5c and Figure 5d, respectively. For yield strength, uncertainty in the effect of the test temperature was significantly greater than any of the other features in the model. This can be rationalized from the fact that, in the temperature-dependent yield strength model, the test temperature is by far the most important feature. Nonetheless, variance in the temperature effect only accounts for 18% of the observed variance amongst the entire feature set. Moreover, some features (e.g., Hf and Al element fractions) which lie completely outside of the ten most important features in Figure 5a are among the major contributors to variance in the model output as determined from SHAP. A similar phenomenon is seen when performing the same analysis on the room-temperature plasticity model, where the feature which contributes most to the variance (the annealing time, in hours) is surprisingly not present within the ten most important features. In this case, however, while the annealing time did not appear among the most important features, two related quantities, annealing temperature and "Anneal", a one-hot encoded variable indicating if the alloy had undergone annealing, were deemed to be highly important through SHAP. In both the cases of test temperature in the yield strength model and the annealing time in the plasticity model, the variance in the feature effects on the model output are believed to partly result from the sparsity of the dataset, where the data is heavily skewed towards room-temperature measurements of yield strength and lacks sufficient data to confidently determine dependence of the plasticity on the annealing conditions. Future experimental efforts will serve to bolster the currently sparse data on these experimental conditions to improve uncertainty in the effects of each of these features. Nevertheless, the current model is capable of highlighting the interplay between certain key physics-based microstructural properties and their effect on both the yield strength and plasticity. Physics-based theories describe the solid solution strengthening as proportional to $F^{4/3}x^{2/3}$ where $F$ is the interaction force and $x$ is the molar solute concentration. The interaction force is, in turn, proportional to $\beta\delta_r + \delta_G$, where $\delta_r$ and $\delta_G$ are the lattice distortion and modulus distortion, and $\beta$ is a constant, which is related with the difference in the interaction forces between screw and edge dislocations and the local stress field caused by solute atoms. In Figure 5e and Figure 5f, these theoretical relationships (captured through the calculation of $\beta\delta_{r(G)}^{4/3}x^{2/3}$) are shown with the elemental distortion contributions to yield strength and plasticity which appeared in the feature importance plots of Figure 5a and Figure 5b (i.e., $\delta G_{\mathrm{Ta}}$, $\delta r_{\mathrm{Hf}}$, $\delta G_{\mathrm{Mo}}$, and $\delta r_{\mathrm{V}}$). In agreement with solid solution strengthening theory, larger distortion generally leads to a positive impact on the yield strength determined from SHAP analysis. On the other hand, increased distortion contributions from key constituent elements frequently leads to a decrease in plasticity, thus highlighting one origin of the inherent trade-off between the two mechanical properties. However, in some cases, most notably for vanadium, increase in distortion can lead to both a positive impact in yield strength *and* plasticity. This observation of strength-plasticity synergy was experimentally observed by Wang *et al.* for increasing vanadium concentration in the $V_xNbMoTa$ alloy system[34]. Our model suggests that this possibility for strength-plasticity synergy may extend more generally into a wide RHEA composition space.

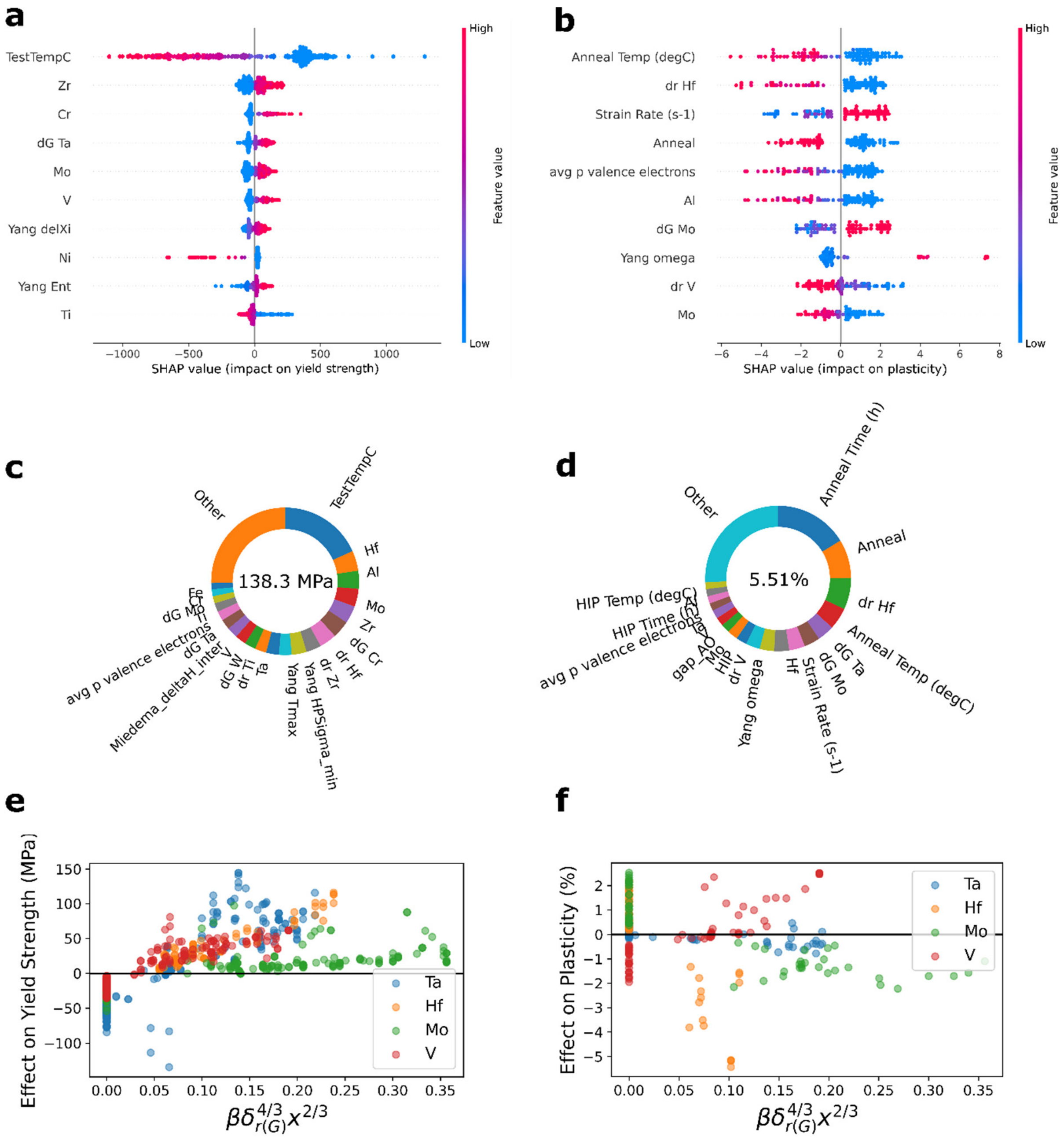

**Figure 5.** Model interpretation through computing Shapley values. Summary of the SHAP values of the ensemble yield strength model (a) and the ensemble plasticity model (b). Negative SHAP values correspond to observations (i.e., individual members of the RHEA database) where the feature had a negative effect on the output (yield strength or plasticity), whereas positive SHAP values correspond to observations where the feature had a positive impact. Red points indicate observations where the feature had a high value, relative to other observations in the dataset, whereas blue points indicate observations where the feature had a low value. The twenty most important features (as determined by their mean absolute SHAP value) are shown on the left in descending order of importance. (c) and (d) show the contribution of individual features to the total uncertainty (shown in the center of the circle) of the feature effects determined by SHAP.

Features explaining 75% of the uncertainty are shown, with all remaining feature effects indicated by "Other". (e) and (f) show the effect on yield strength and plasticity, respectively, of the elemental lattice and modulus distortion-related terms identified as key contributing features in panels (a) and (b). β was assumed to be equal to 9 (indicating a mixture of screw and edge dislocations) in accordance with Lee *et al.*[35] The vertical cluster of points at zero distortion correspond to alloys where the given element was not present.

**Discussion**

Due to the small datasets which are typical within the materials science community, it is imperative that alternative forms of statistical validation be leveraged in order to improve confidence in uncertainty estimates. To this end, while experimental synthesis and characterization remains the gold standard for validation, statistical validation techniques which properly represents unseen data and accounts for epistemic error (e.g., repeated *k*-fold cross-validation) should not be undervalued as offering an appropriate alternative to assessing model generalizability. The presence of repeated or correlated data in the original dataset, however, can cause the generalization error to be underestimated. We have demonstrated the use of ensemble models which provides for inherent uncertainty quantification. The framework is built upon existing theoretical knowledge in microstructural mechanics and dislocation theory, thereby enabling the model to be interpreted and compared against physical intuition. We envision that the framework we have outlined herein can be extended to many materials science problems outside of high-entropy alloys. While outside of the scope of this work, our findings have laid the groundwork for high-throughput screening and simultaneous optimization of yield strength and plasticity. This will be the subject of a future publication.

**Methodology**

The neural network models were implemented in TensorFlow[36] using the Keras[37] subroutine. A range of model architectures and hyperparameters were tested. To optimize the architecture and corresponding hyperparameters, the Optuna[38] Python package was used. The following ranges of architecture and hyperparameters were chosen to bound the optimization. To perform the optimization, Optuna uses a sampling and pruning algorithm to converge to an optimal set of hyperparameters. In this work, the tree-structured Parzen estimator[39], a Bayesian optimization algorithm, and a pruning algorithm which removed undefined objective function values were employed for sampling and pruning, respectively.

| Parameter | Range in Trials |
|---|---|
| Neurons in hidden layers | 4 – 256 |
| Number of hidden layers | 1 – 3 |
| Number of training epochs | 10 – 500 |
| L2 regularization weight decay | $10^{-10}$ – $10^{-3}$ |
| Optimizer | Adam[40] or SGD[41] |
| Learning rate (Adam) | $10^{-4}$ – $10^{-1}$ |
| Learning rate (SGD) | $10^{-5}$ – $10^{-1}$ |
| Momentum (SGD) | $10^{-4}$ – $10^{-1}$ |

Using the RHEA compositions contained in the yield strength and plasticity datasets, a set of composition-based features were calculated on the basis of atomic properties. These composition-based features included such quantities as RHEA density, melting point, atomic size mismatch, mixing enthalpy, configurational entropy, etc. Among these composition-based features are functionalities we have added to the Matminer code to allow for the computation of theoretical quantities known to control and/or describe strengthening mechanisms. These include: lattice distortion, modulus distortion, Hall-Petch parameters, Poisson ratio, etc.

Some noteworthy descriptors that bear discussion are ones that have been traditionally used for interpreting phases, such as the atomic size mismatch ($\delta$), the enthalpy of mixing ($\Delta H_{\mathrm{mix}}$), the entropy of mixing ($\Delta S_{\mathrm{mix}}$), $\Omega$, and $\Phi$. The equations defining these five descriptors are given as[42,43],

$$\delta = \sqrt{\sum_{i=1}^{n} x_i (1 - r_i/\bar{r})^2} \tag{1}$$

$$\Delta H_{\mathrm{mix}} = \sum_{i=1, i \neq j}^{n} \Omega_{ij} x_i x_j \tag{2}$$

$$\Delta S_{\mathrm{mix}} = -R \sum_{i=1, i \neq j}^{n} x_i \ln x_j \tag{3}$$

$$\Omega = \frac{T_{\mathrm{m}} \Delta S_{\mathrm{mix}}}{|\Delta H_{\mathrm{mix}}|} \tag{4}$$

$$\Phi = \left| \frac{\Delta H_{\mathrm{mix}} - T_{\mathrm{m}} \Delta S_{\mathrm{mix}}}{\Delta H_{\mathrm{max}} - T_{\mathrm{max}} \Delta S_{\mathrm{max}}} \right| \tag{5}$$

In Equations 1 – 5, $i$ and $j$ represent the $i^{\mathrm{th}}$ and $j^{\mathrm{th}}$ element, $r_i$ is the atomic radius of the $i^{\mathrm{th}}$ element, $\bar{r}$ is the average atomic radius of the alloy, $\Omega_{ij} = 4\Delta H_{\mathrm{AB}}^{\mathrm{mix}}$ (where $\Delta H_{\mathrm{AB}}^{\mathrm{mix}}$ is the enthalpy of mixing of binary alloys), $x$ is the element fraction, $T_{\mathrm{m}}$ is melting temperature estimated from a rule-of-mixtures, $\Delta H_{\mathrm{max}}$ is the maximum enthalpy of mixing of a binary combination of the elements in the alloy, $T_{\mathrm{max}}$ is the maximum melting temperature of a single element in the alloy, and $\Delta S_{\mathrm{max}}$ is the maximum entropy of mixing of a binary alloy (i.e., an equiatomic binary alloy).

Recently, lattice distortion and moduli of distortion are thought to be important parameters for determining phase formation and mechanical properties of HEAs[44,45]. Therefore, we have included them in our descriptor calculation. The lattice distortion around atom $i$ can be defined according to Senkov *et al.*[46] as,

$$\delta_{ai} = \frac{9}{8}\sum_{j} x_j \delta_{aij} \tag{6}$$

The 9 in the numerator is due to total number of atoms in the $i$ -centered cluster in the BCC lattice, while the 8 in the denominator is due to the number of atoms around $i$ in the cluster (excluding $i$). The reduced atomic size difference, $\delta_{aij}$, is defined as,

$$\delta_{aij} = \frac{2(r_i - r_j)}{(r_i + r_j)} \tag{7}$$

Similarly, the modulus of distortion, $\delta_{Gi}$, is the defined as,

$$\delta_{Gi} = \frac{9}{8}\sum_{j} c_j \delta_{Gij} \tag{8}$$

$$\delta_{Gij} = {2(G_i - G_j)}/{(G_i + G_j)} \tag{9}$$

where $G$ is the shear modulus. The atomic radii and moduli were collected from published values[42,47].

In addition to solid solution strengthening, grain boundary strengthening is another mechanism through which the mechanical properties of an alloy can be affected. The primary empirical equation governing the contribution of grain boundary strengthening to the observed mechanical property is the Hall-Petch relationship[48,49]. The Hall-Petch relationship is frequently expressed as,

$$\sigma_{\mathrm{HP}} = \sigma_0 + Kd^{-\frac{1}{2}} \tag{10}$$

Where $\sigma_0$ is the base strength of the material, $K$ is the locking parameter, and $d$ is the grain size of the material. In alloys, the base strength is typically derived from the the individual element yield strengths using the rule-of-mixtures. The locking parameters of all elements of interest in this study have been tabulated in the literature[50]. Here, a rule-of-mixtures has been used to estimate the locking parameter of each alloy based on their elemental composition. The grain size, $d$, is closely related to the specific processing conditions affecting grain growth kinetics[51,52], and is typically not reported in the HEA literature. However, some HEA literature have performed detailed studies on the effect of ubiquitously reported processing conditions (e.g., annealing temperature, annealing time, etc.) on the grain size. Therefore, the specific processing conditions which we have collected from the original literature cited by Couzinie *et al.*[16] serve as a surrogate for a more detailed microstructural knowledge.

*Experimental Methods*

Ingots of AlMoTaTiZr and $Al_{0.239}Mo_{0.123}Ta_{0.095}Ti_{0.342}Zr_{0.201}$ were prepared by melting high-purity elements (all element purity > 99.5 at.%) using an argon-backfilled arc melter at room temperature. All elemental components were measured to an accuracy of $\pm 10^{-3}$ g. The ingots were cast and then cut into cylindrical geometries [Diameter: 4 mm, Length: 6.5 - 8 mm, Aspect ratio: 1.5-2] for both room- and high-temperature (900 °C) compression testing, employing a Materials Testing System (MTS) attached with a 647 Hydraulic Wedge Grip and an MTS 653 Furnace. The strain rate was $10^{-3}$ $s^{-1}$ for compression experiments. Lastly, prior to compression tests, all samples were polished using SiC polishing paper to 1200 grit prior to testing.

**Acknowledgments**

This work was funded by the Office of Naval Research of the United States under the Small Business Technology Transfer program (Contract # N68335-20-C-0402).

**Data Availability**

The data used to generate the machine learning models (compositions, feature values, measured yield strength, plasticity, etc.) are included in the SI. A separate file containing the recent validation data that was collected for the yield strength and plasticity models is also included in the SI.

**Code Availability**

Readers are requested to contact the authors.

**Supplemental Information**

**Yield Strength-Plasticity Trade-off and Uncertainty Quantification for Machine-learning-based Design of Refractory High-Entropy Alloys**

Stephen A. Giles[1], Hugh Shortt[2], Peter K. Liaw[2], Debasis Sengupta[1]

[1] CFD Research Corporation, 6820 Moquin Drive NW, Huntsville, AL 35806

[2] Department of Materials Science and Engineering, The University of Tennessee, Knoxville, TN 37996.

## SUPPLEMENTARY INFORMATION

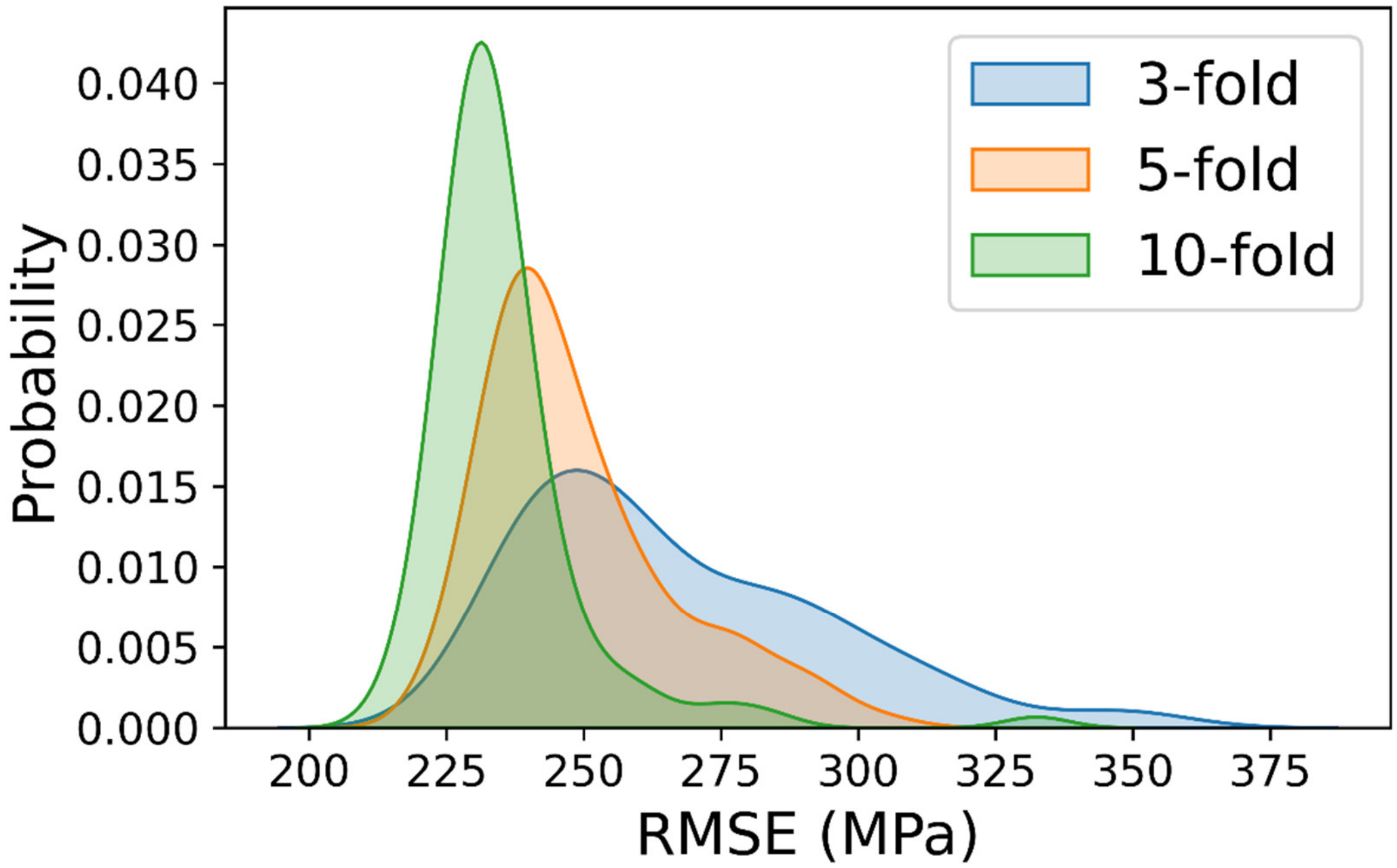

**Figure S6.** Distribution of error in yield strength prediction, depicted by kernel density estimate (KDE), for ReLU activation as a function of number of folds.

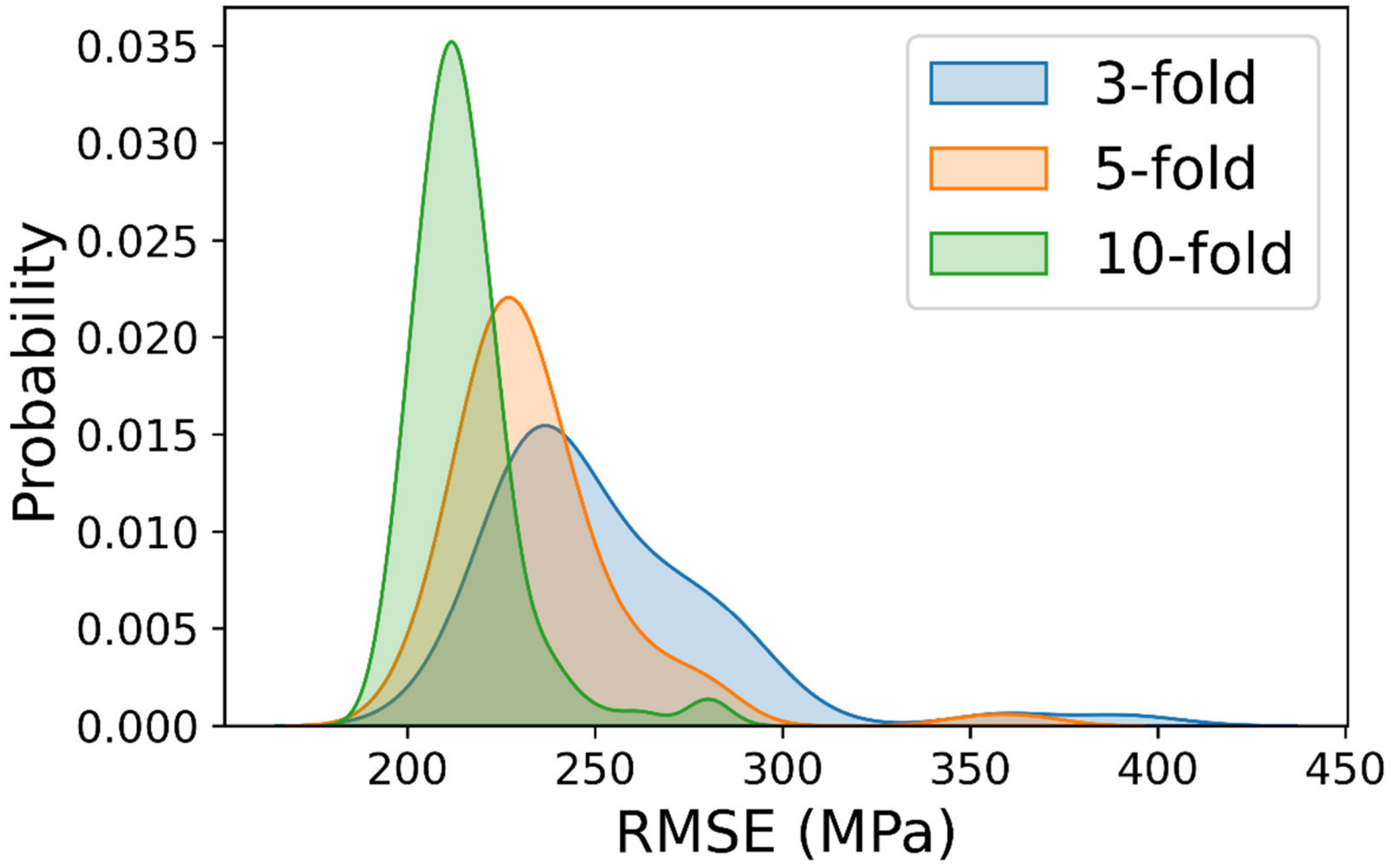

**Figure S7.** Distribution of error in yield strength prediction, depicted by KDE, for SELU activation as a function of number of folds.

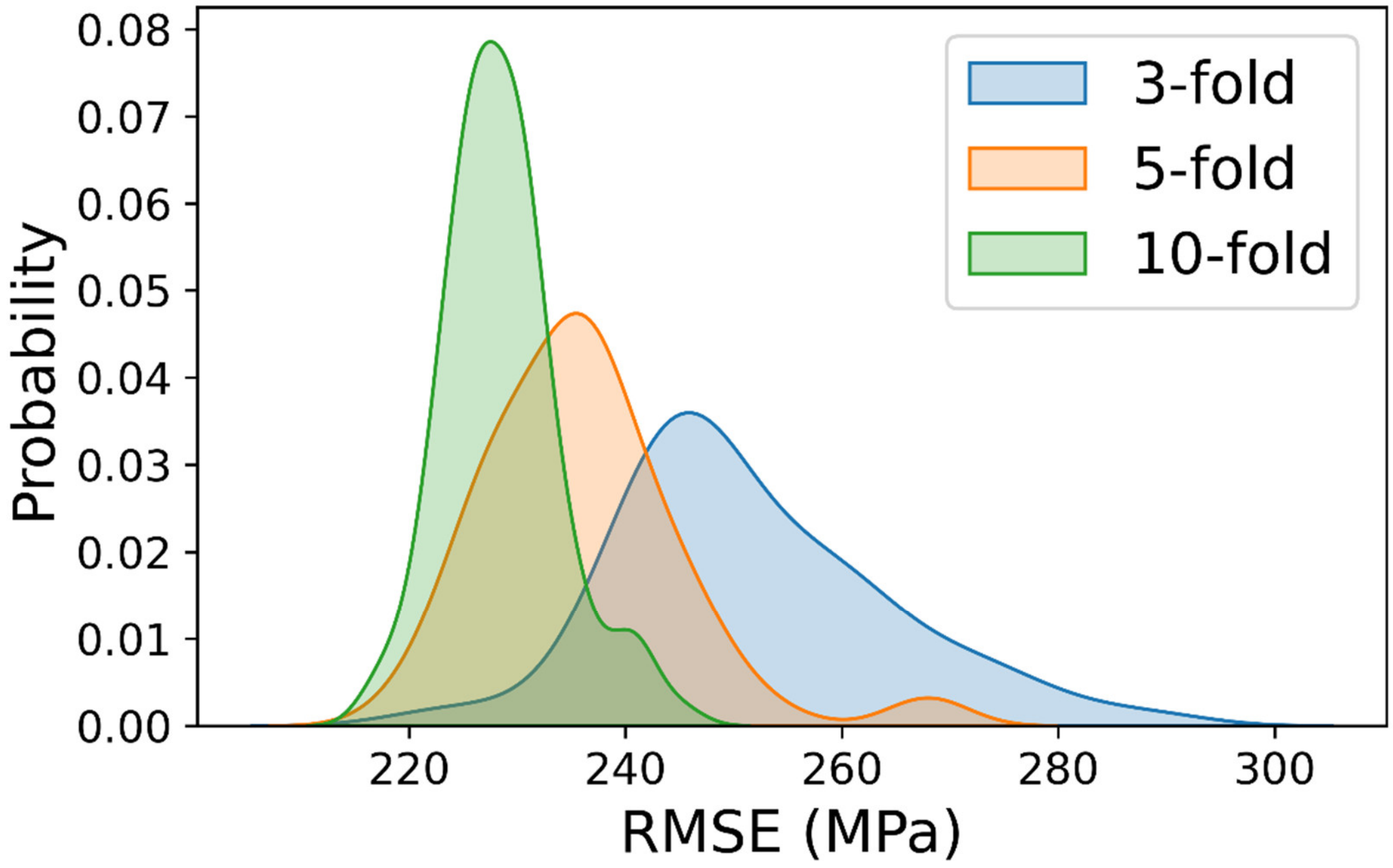

**Figure S8.** Distribution of error in yield strength prediction, depicted by KDE, for the random forest model as a function of number of folds.

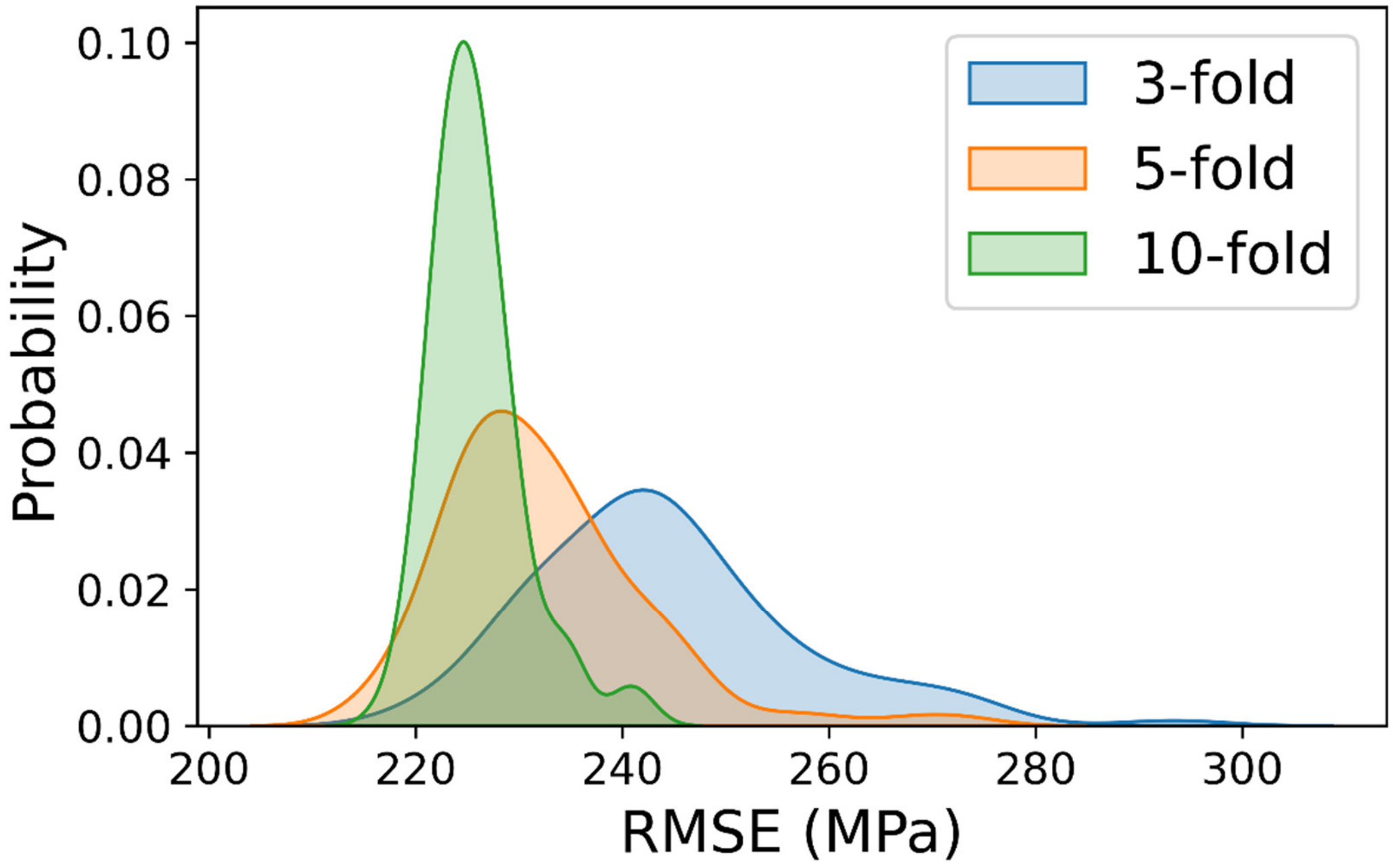

**Figure S9**. Distribution of error in yield strength prediction, depicted by KDE, for the gradient boosting model as a function of number of folds.

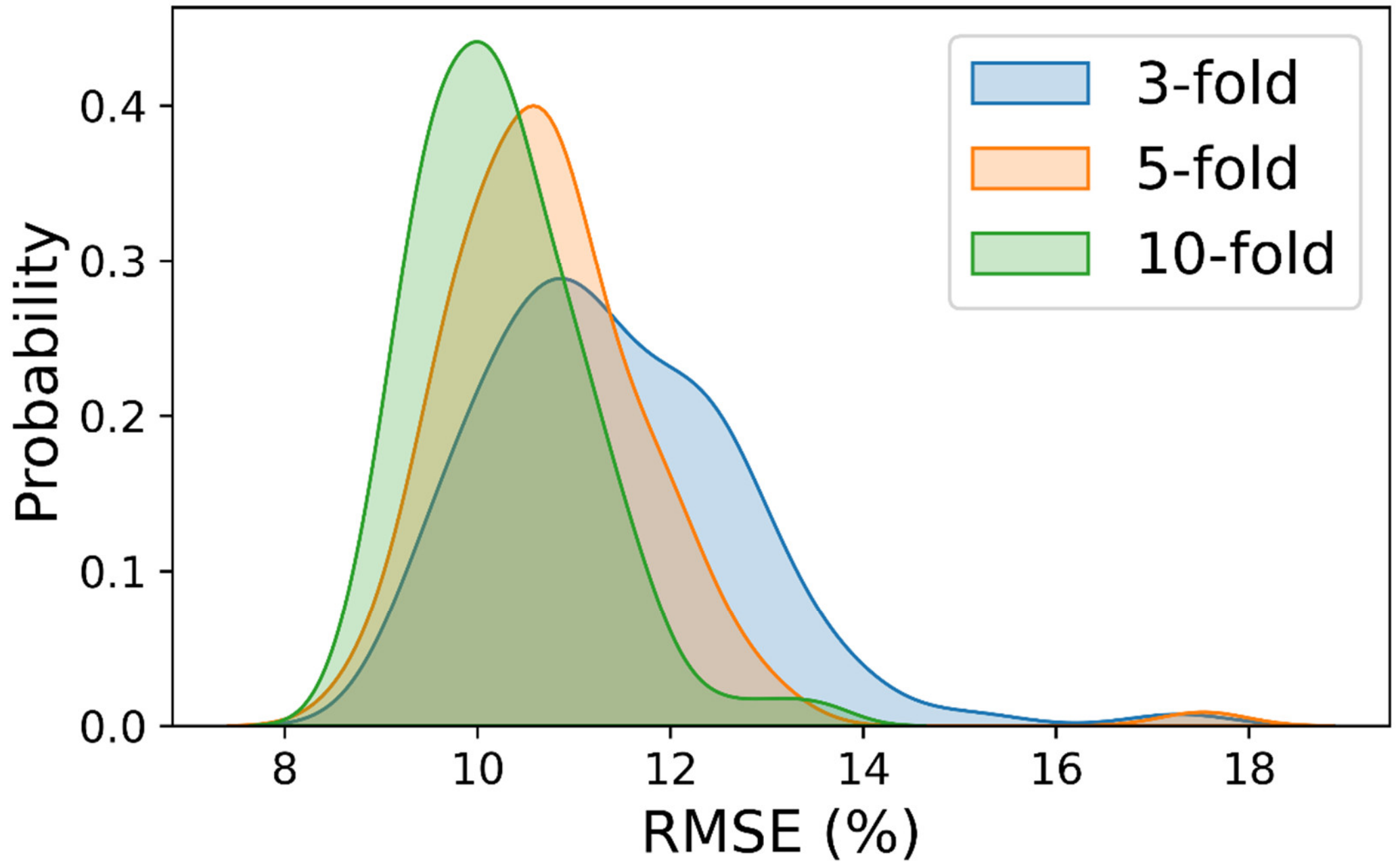

**Figure S10.** Distribution of error in plasticity prediction, depicted by KDE, for the neural network model with ReLU activation as a function of number of folds.

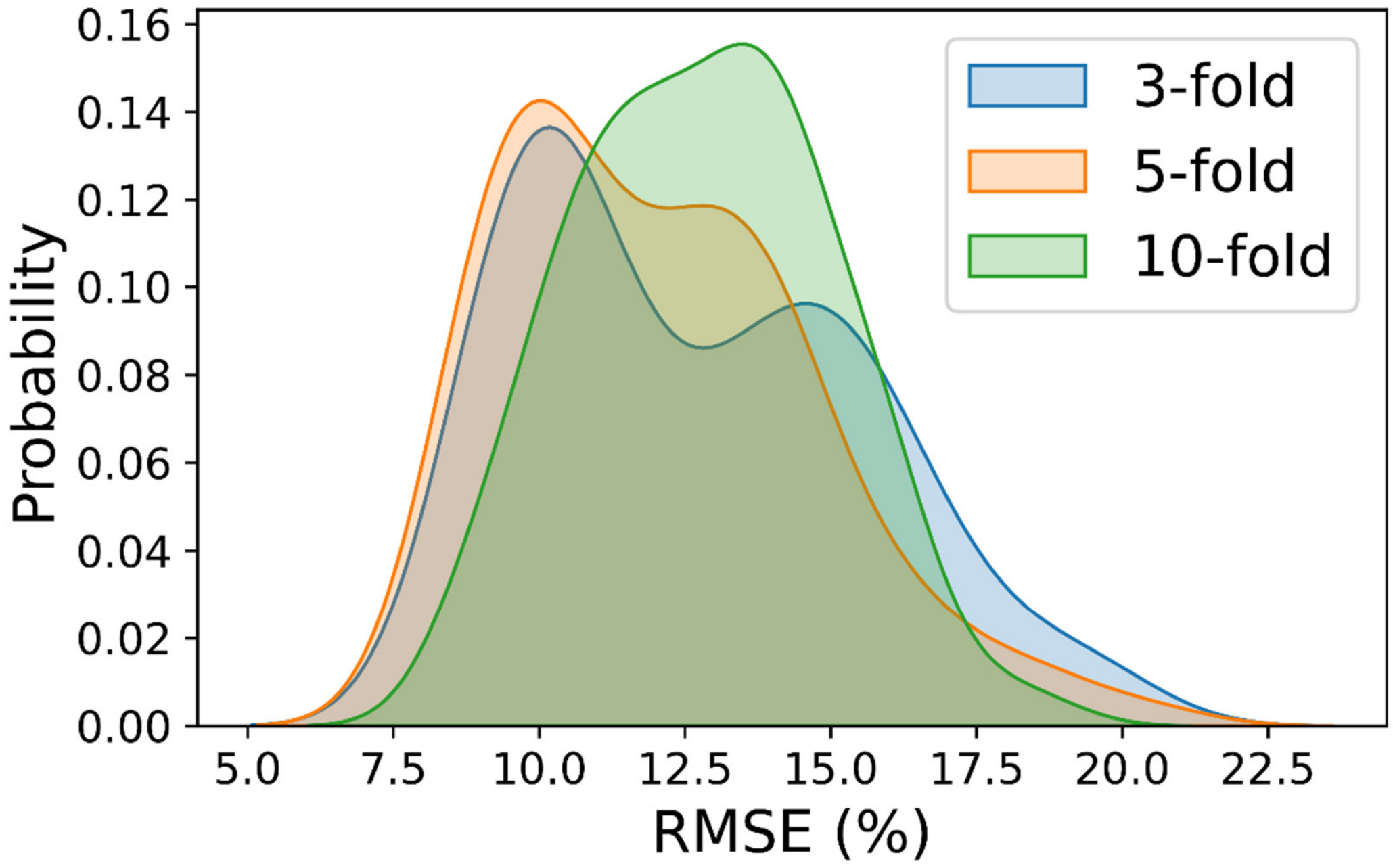

**Figure S11.** Distribution of error in plasticity prediction, depicted by KDE, for the neural network model with SELU activation as a function of number of folds.

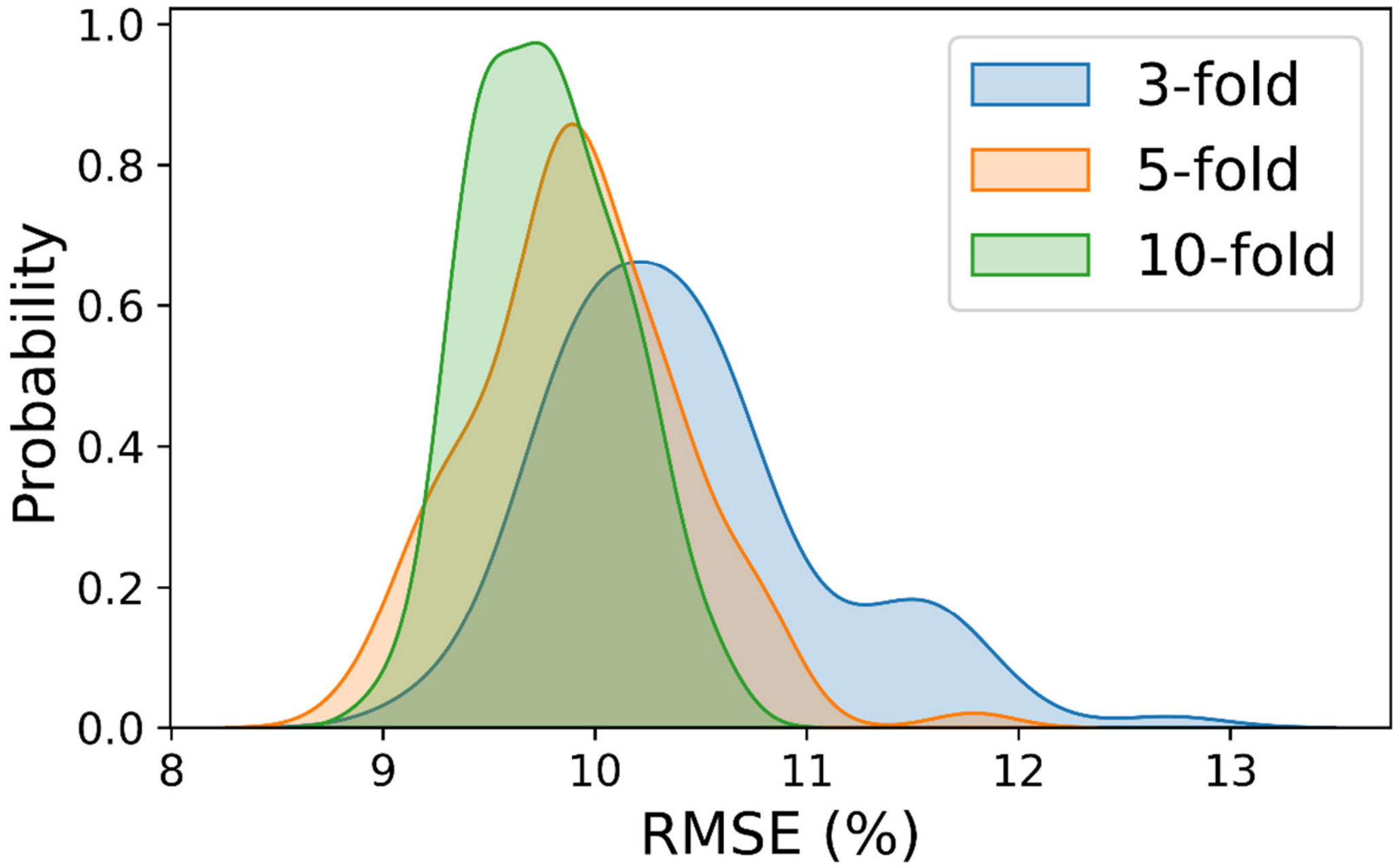

**Figure S12.** Distribution of error in plasticity prediction, depicted by KDE, for the random forest model as a function of number of folds.

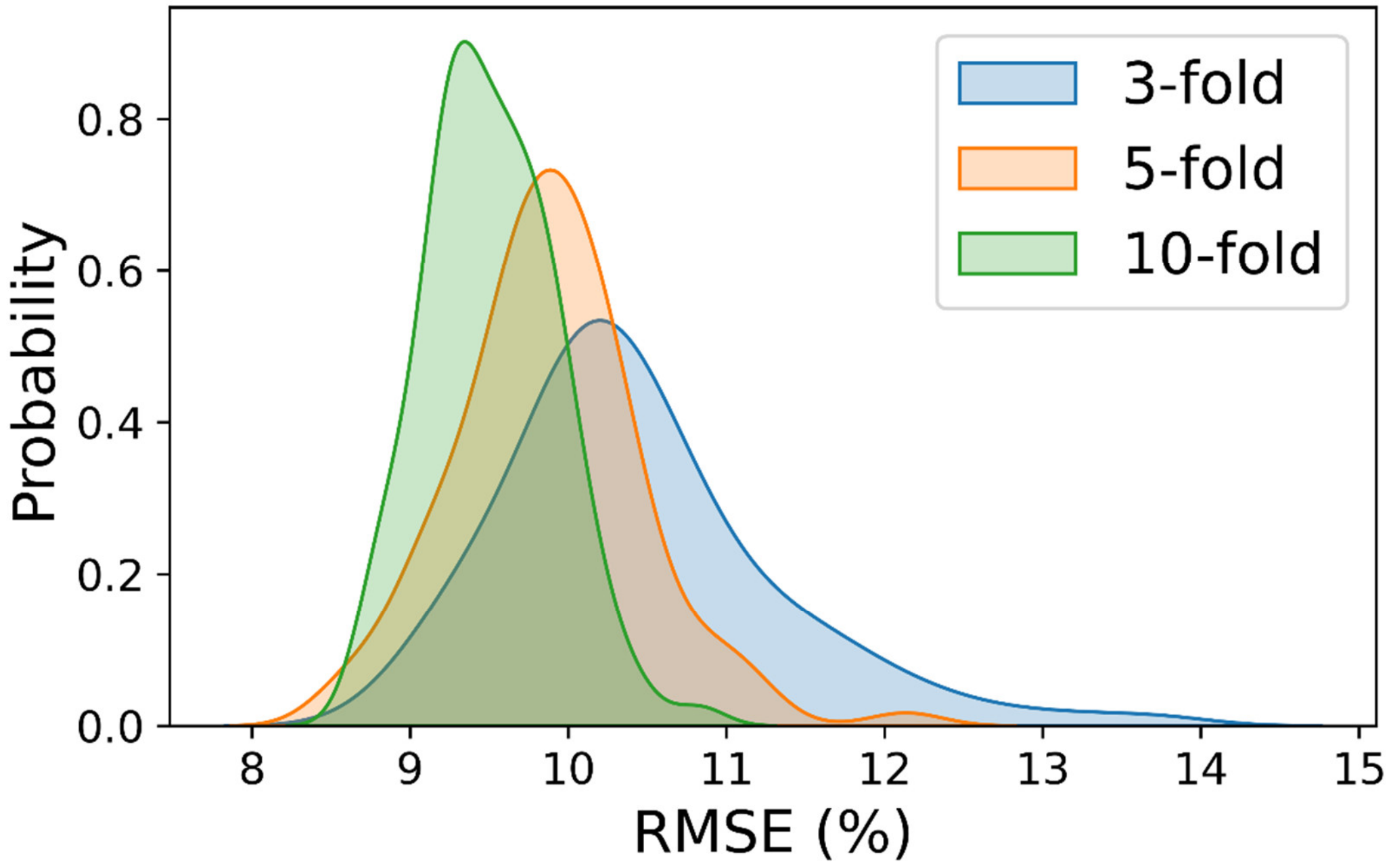

**Figure S13.** Distribution of error in plasticity prediction, depicted by KDE, for the gradient boosting model as a function of number of folds.

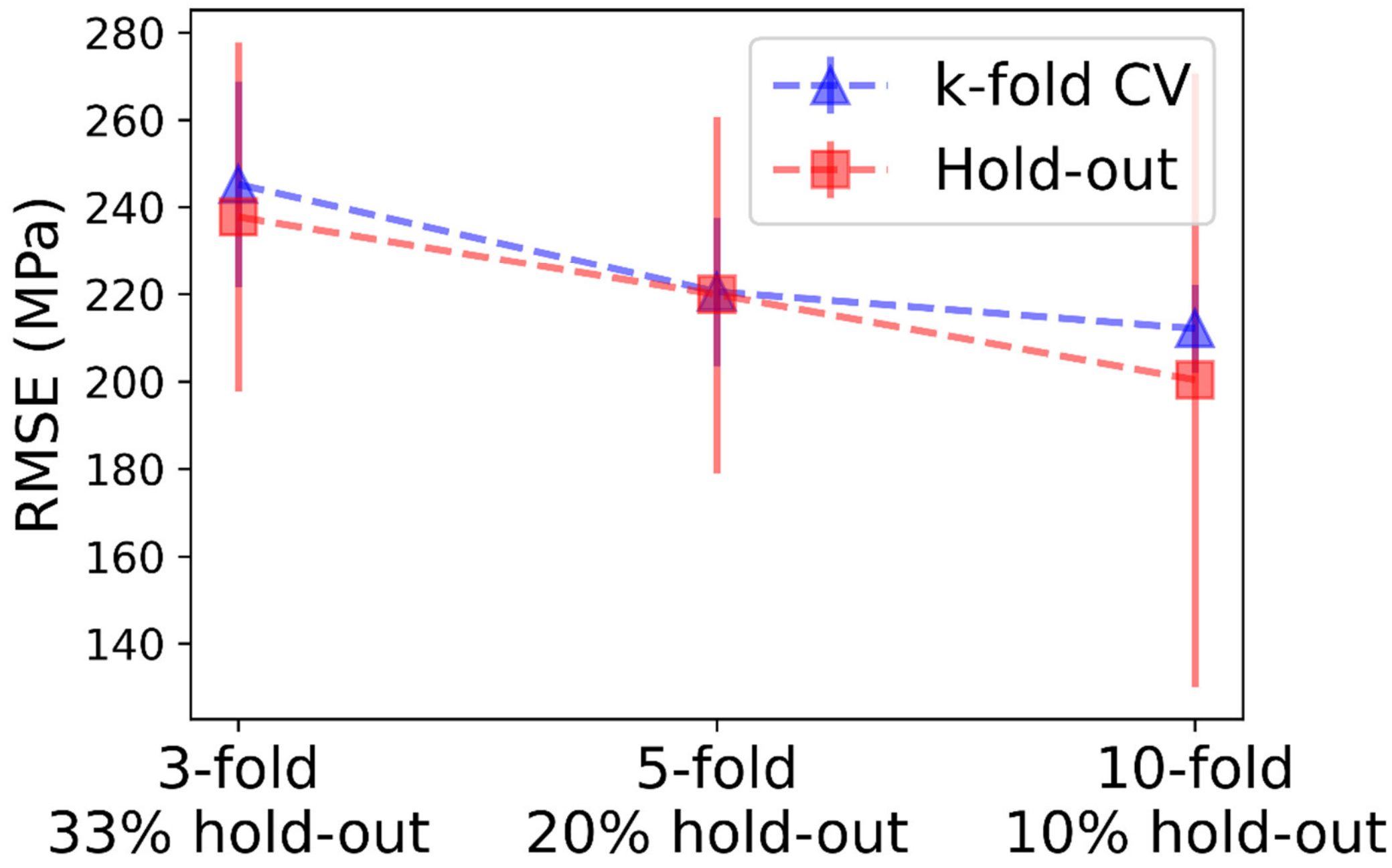

**Figure S14.** Comparison of *k*-fold cross-validation error estimates to their corresponding hold-out error estimates, where a portion of the dataset has been entirely withheld.

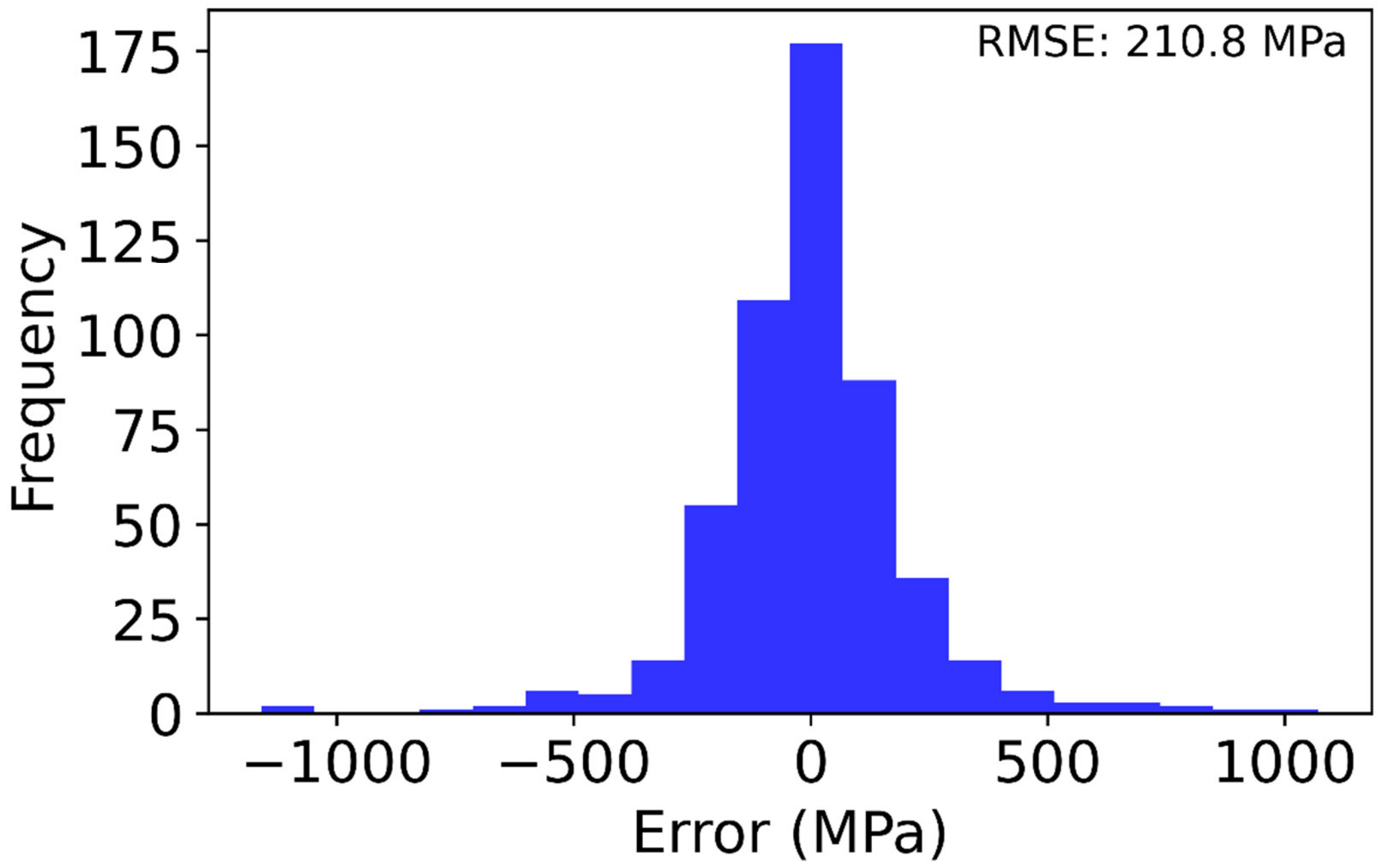

**Figure S15.** Example error distribution of yield strength for the neural network model with Leaky ReLU activation. Errors greater than zero correspond to an overprediction of the true values, whereas errors less than zero correspond to an underprediction.

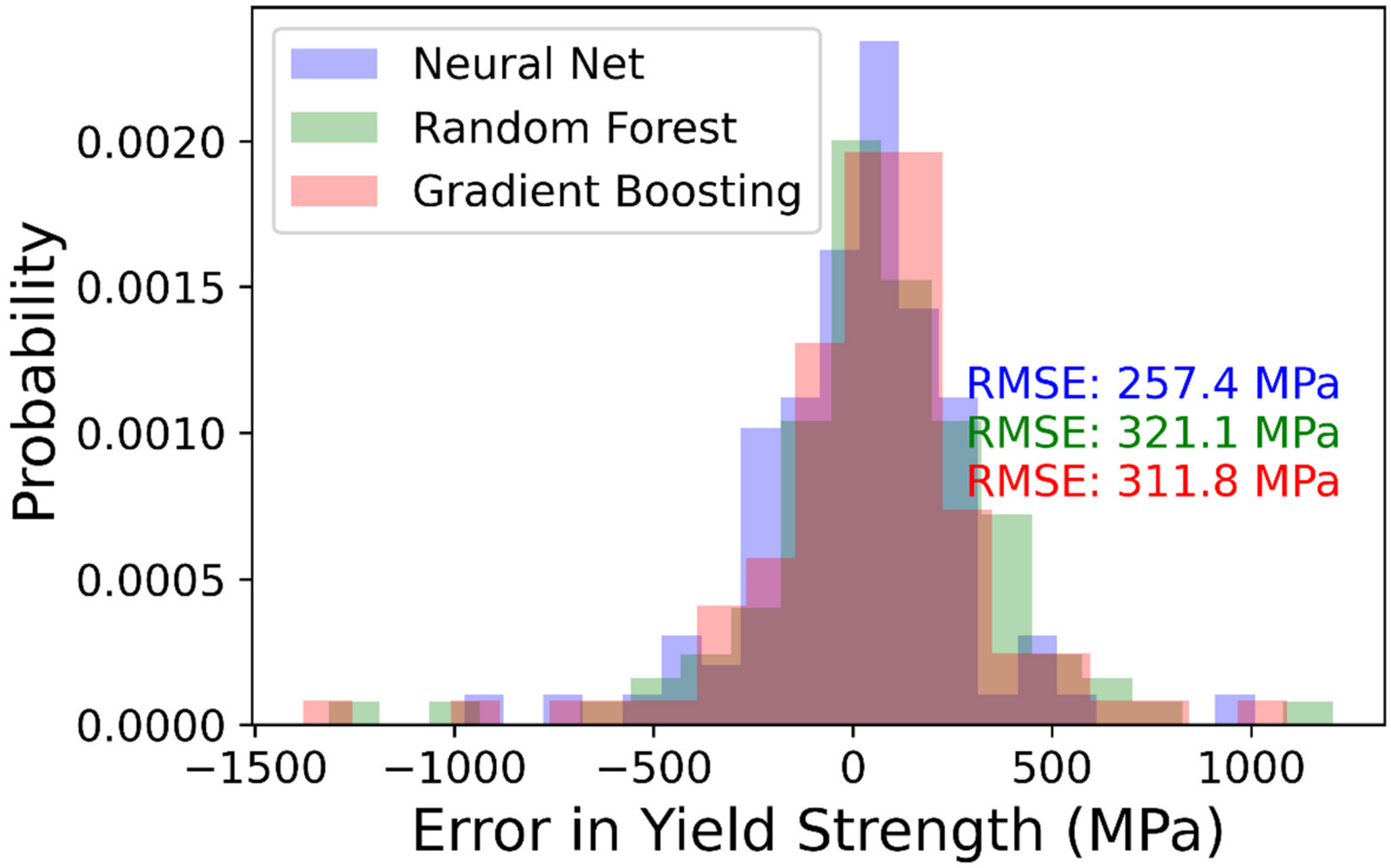

**Figure S16.** Distribution of the 10-fold cross-validation error in the prediction of unique alloy composition yield strengths for each of the three model types, where the Neural Net results used Leaky ReLU activation.